# 21 cm Intensity Mapping

Jeffrey B. Peterson[1], Roy Aleksan[2], Réza Ansari[3], Kevin Bandura[1], Dick Bond[4], John Bunton[5], Kermit Carlson[6], Tzu-Ching Chang[4,7], Fritz DeJongh[3], Matt Dobbs[8], Scott Dodelson[6], Hassane Darhmaoui[9], Nick Gnedin[6], Mark Halpern[10], Craig Hogan[6], Jean-Marc Le Goff[2], Tiehui Ted Liu[6], Ahmed Legrouri[9], Avi Loeb[11] ,Khalid Loudiyi[9], Christophe Magneville[2], John Marriner[6], David P. McGinnis[6], Bruce McWilliams[1], Marc Moniez[3], Nathalie Palanque-Delabruille[2], Ralph J. Pasquinelli[6], Ue-Li Pen[4], Jim Rich[2], Vic Scarpine[6], Hee-Jong Seo[6], Kris Sigurdson[10], Uros Seljak[12], Albert Stebbins[6], Jason H. Steffen[6], Chris Stoughton[6], Peter T. Timbie[13], Alberto Vallinotto[6], Stuart Wyithe[14] and Christophe Yeche[2]

[1] Dept. of Physics, Carnegie Mellon University
[2] SPP-IRFU, CEA-Saclay
[3] LAL-Orsay
[4] Canadian Institute for Theoretical Astrophysics
[5] CSIRO
[6] Fermi National Accelerator Laboratory
[7] Academica Sinica Institute of Astronomy and Astrophysics
[8] McGill University
[9] School of Science & Engineering, Al Akhawayn University in Ifrane, Morocco
[10] University of British Columbia
[11] Dept. of Astronomy, Harvard University
[12] University of California, Berkeley
[13] University of Wisconsin
[14] University of Melbourne

email contact: jbp@cmu.edu

**21-cm Intensity Mapping.**

Using the 21 cm line, observed all-sky and across the redshift range from 0 to 5, the large scale structure of the Universe can be mapped in three dimensions. This can be accomplished by studying specific intensity with resolution ~ 10 Mpc, rather than via the usual galaxy redshift survey. The data set can be analyzed to determine Baryon Acoustic Oscillation wavelengths, in order to address the question **What is the nature of Dark Energy?** In addition, the study of Large Scale Structure across this range addresses the questions **How does Gravity effect very large objects?** and **What is the composition our Universe?** The same data set can be used to search for and catalog time variable and transient radio sources.

## Three dimensional 'Intensity Mapping' of the Universe.

Previously, the primary method to obtain a 3-D map of large scale structure in the universe has been via galaxy redshift surveys, that is by isolating millions of individual galaxies, recording spectra for each and determining each redshift. This has been done primarily using optical spectroscopy. Combining all optical surveys, about 1% of the observable universe has been mapped this way.

Several authors [4, 6, 9, 10] have proposed a new strategy to economically map much larger volumes of the universe, by measuring the collective emission of many galaxies without individual detections. We call this technique "Intensity Mapping". Under this scheme we use a telescope with resolution and sensitivity sufficient to measure the Large Scale Structure, and especially the Baryon Acoustic Oscillation wavelengths, but without the resolution or sensitivity to detect individual galaxies. Roughly 50 times as much collecting area would be needed to carry out a 21 cm galaxy redshift survey of the same volume.

The 21-cm transition of hydrogen is the dominant spectral line in astronomical observations at frequencies less then 1420 MHz and is an isolated transition. This allows direct translation of the intensity versus frequency to HI density versus redshift.

The technique we propose is similar to the velocity mapping of galaxy disks using HI or the mapping of Galactic molecular clouds using CO line emission, however here we suggest carrying out such observations across the entire sky, and through a cosmological span of redshift.

## Measuring Baryon Acoustic Oscillations with the 21 cm line of hydrogen

We propose to study the expansion history using the Baryon Acoustic Oscillations (BAO) method. During the first 380,000 years of the expansion, cosmic matter was highly ionized and this allowed acoustic waves to propagate. At the end of the ionized era, acoustic waves no longer propagated, but any wave crests present at that time left behind density enhancements. Because these imprints were put in place throughout the universe, everywhere with the same physical size, they can serve as a standard ruler. [1, 3, 5, 8] A 21-cm intensity-mapping telescope can be used to look at this 'ruler' across a range of redshifts, and by measuring the angular and redshift space BAO wavelength at each redshift it is possible to trace out the history of the cosmic expansion. By making this measurement across the redshift range where dark energy kicks in, such data sharply constrain dark energy models.

The BAO correlation length is enormous: 150 Mpc is a few percent of the current Hubble length. This large scale presents both an advantage and a challenge. On small scales, the process of non-linear gravitational collapse wipes out primordial structure. Here the large size of BAO features is an advantage: it guarantees the BAO signal is not erased. Non-linear processes do slightly shift the BAO correlation length, but this shift can be readily calculated or measured and the necessary small corrections can be made. Because of this the BAO method is considered by many to be the cleanest of the proposed methods to measure the properties of dark energy. The challenge presented by the large correlation length is that any BAO dark energy experiment must examine a huge co-moving volume. The Intensity-mapping scheme aims to provide the necessary volume.

Simulations indicate the 21 cm intensity mapping technique will allow very precise constraints to the dark energy equation of state [4]. The sensitivity comes from the large volume of the survey. We will learn in detail how the dark energy negative pressure (its anti-gravity action) evolved as the universe expanded. Results of these simulations showing the sensitivity to dark energy models[2] are summarized in figure 1.

As explained below, to succeed in our 21 cm BAO observations we must separate 21 cm line emissions from the much stronger broadband emission by the our Galaxy and by many extragalactic sources. This is the most challenging part of the program.

**Study of Large Scale Structure.**

Because 21-cm intensity mapping experiments can cover a huge co-moving volume, such data can provide unprecedented precision in the 3D measurement of large-scale structure. This may allow tight constraints to cosmological parameters [6][10][11] while also testing the standard gravitational collapse paradigm.

The explanation for the current universal acceleration may lie not in an unseen dark-energy constituent but instead may require a revised theory of gravitation. To provide observational constraints on gravitation theories one needs data not just on the history of the Hubble parameter but on finer scales as well. Figure 2 shows a simulation of the 3-D 21 cm intensity structure we expect to detect.

## 21 cm Signal Levels

There are similarities between 21 cm intensity-mapping observations and CMB observations, but also significant differences. In terms of brightness temperature the BAO signal is smaller for 21cm than the first acoustic peak in the CMB (20μK vs. 100μK). However, the 21 cm structure is three-dimensional while the CMB sky is two dimensional, allowing an accumulation of BAO signal over substantial spans of redshift. The contamination from Galactic and extra-galactic emission (collectively known as "foregrounds") is much greater for 21 cm observations. Removal of this interfering emission is the key difficulty for 21 cm intensity-mapping observations and this is discussed below.

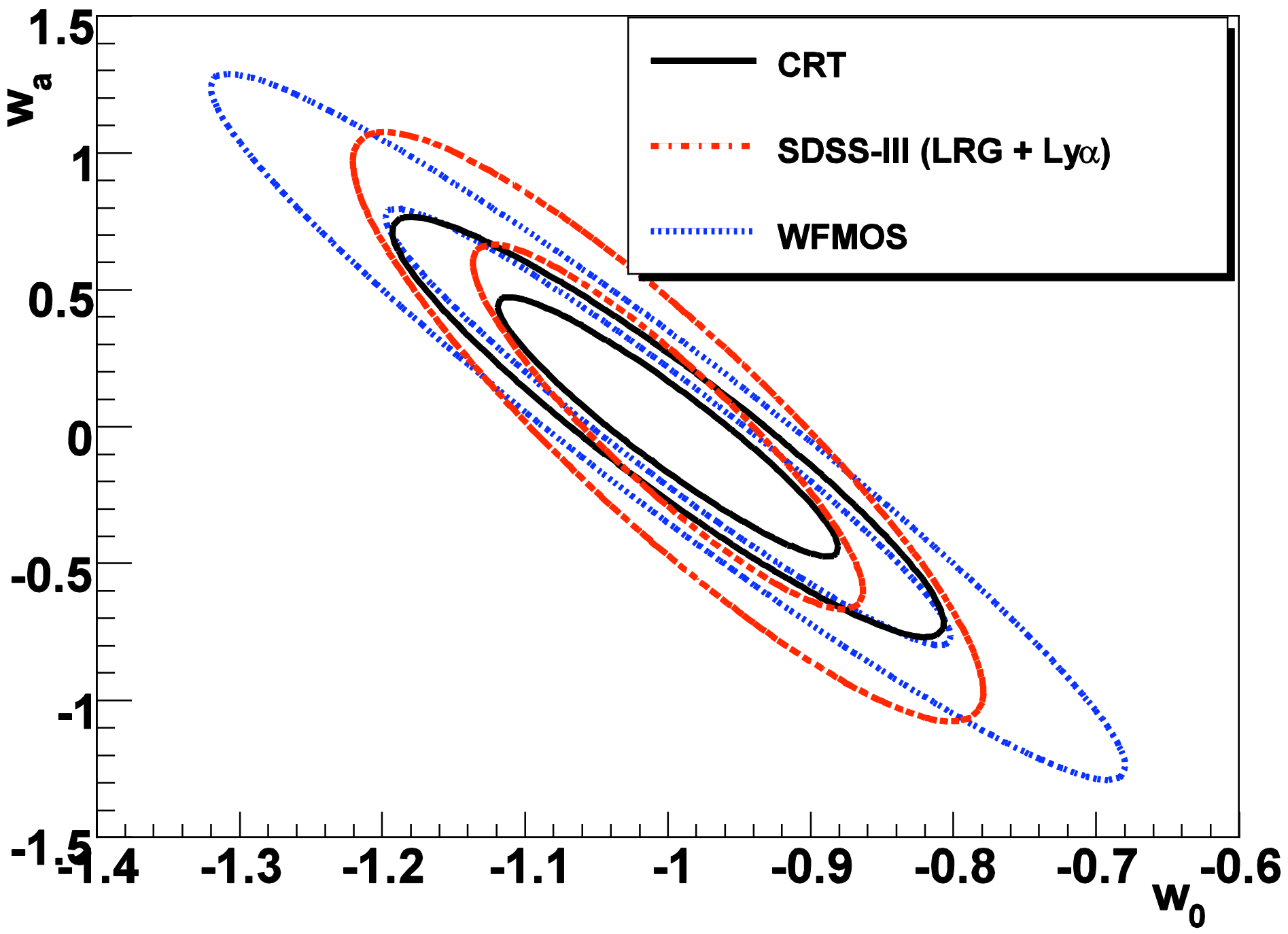

**Figure 1. Sensitivity to Dark Energy Models. The plot shows projected error ellipses in the parameter space laid out by the Dark Energy Task Force. $w_0 = p/\rho$ is the dark energy equation of state, and $w_a$ is the first derivative of w. Plotted in black are 1-σ and 2-σ contours for HI Intensity Mapping assuming a Cylinder Radio Telescope of area 10,000 sq meters. Also plotted are contours for the optical BAO experiments SDSS-III after combining the Large Red Galaxy and Lyman-alpha surveys (red) and WFMOS (blue).**

The strategy for removal of astronomical foreground sources is to subtract their smooth continuum spectra as discussed below. However, even if we had a perfect map of the 21cm intensity, with all foregrounds completely removed, the accuracy with which we can measure the BAO feature is limited by the volume surveyed. The BAO correlation length is 150 Mpc and the accuracy of the BAO measurement increases as the square root of the number of 150 Mpc sized volumes surveyed. Large volumes are essential to any BAO program to achieve accurate measurement of the evolution of dark energy. As illustrated in figure 1, a 21-cm intensity-mapping BAO program can be competitive with other much more costly surveys and this is primarily because to the large volume surveyed.

The 21cm signal due to Large Scale Structure will be approximately 300 μK RMS at the non-linear scale, 8 Mpc. Figure 2 shows the expected structure.

## Foreground Emission

Synchrotron emission is believed to strongly dominate both the Galactic and extragalactic foregrounds at these wavelengths.

The synchrotron radiation foreground has traditionally been assumed to be close to a power law, as this is what one would expect to be produced by a nearly power law distribution of electron cosmic ray energies. One of our collaborators (Stebbins) has recently estimated how smooth the synchrotron foreground spectrum must be, no matter how the electron energies are distributed. Taking the extreme assumption that the electrons are mono-energetic he finds that the spectrum is still smoother by more than ten orders of magnitude than needed for successful subtraction. In other words, the physics of the synchrotron emission process guarantees a lack of spectral features, allowing the 3-D 21 cm intensity signal to stand out.

Despite the smoothness of the sloping synchrotron spectrum, the slope itself can vary from source to source, and the synchrotron sources are not uniformly distributed across the sky. This means the pixel-to-pixel variation of continuum sky brightness exceeds the amplitude of the 21 cm BAO signal by a factor 1000. Very precise separation of these two source types is needed. Fortunately, the spatial characteristic of the BAO signal, a spherical shell in the correlation function at 150 Mpc, is absent in the synchrotron continuum signal. To separate these two signals the plan is to use a combined spatial and spectral filtering technique, via a CMB-style matrix analysis of the data set. Because the spectral and spatial characters of the continuum and BAO sources are so distinct, it should be possible to create a sharp optimal (Wiener) filter and separate the two emission types.

There will be other foreground emission mechanisms, such as free-free emission and Rydberg line emission but these also lack the correlation-shell signature of baryon oscillations and should be separable from the BAO signal.

**Technological Window of Opportunity.**

Several advances of technology over the past decade make the coming decade ripe for dramatic expansion of GHz radio-astronomy. First, technology for Analog to Digital Conversion--followed by Digital Signal Processing, has been developed in multiple radio astronomy laboratories across the world. For bandwidths around 200 MHz this technology is now mature. Second, room temperature low-noise amplifiers are now readily available. Commercial LNA devices, offering noise temperatures below 15 K, now cost just a few dollars. Cryogenically cooled receivers are no longer essential at these frequencies, opening the possibility to distribute receivers widely, rather than concentrating a few receivers in a cryostat.

**Partnership between the Radio-Astronomy and Particle-Physics Communities.**

The expertise needed for the next phase of expansion of GHz radio astronomy is closely matched to the skill set of the experimental particle physics community. For example, low noise GHz amplifiers are used in stochastic cooling systems at accelerators. Also, particle physics detector systems employ analog to digital conversion at GHz sample rates, with thousands of parallel channels. In addition, the handling and analysis of large data sets is an area of expertise for the particle physics community. These are precisely the areas where technical advance will enable the next generation of GHz radio astronomy surveys. In addition to these areas of technological common interests, many of the astro-physics questions addressed by 21 cm sky surveys (e.g. dark energy, gravitation) fall among the areas of fundamental physics that are the focus of the particle physics enterprise. A partnership between the two communities makes sense for both sides.

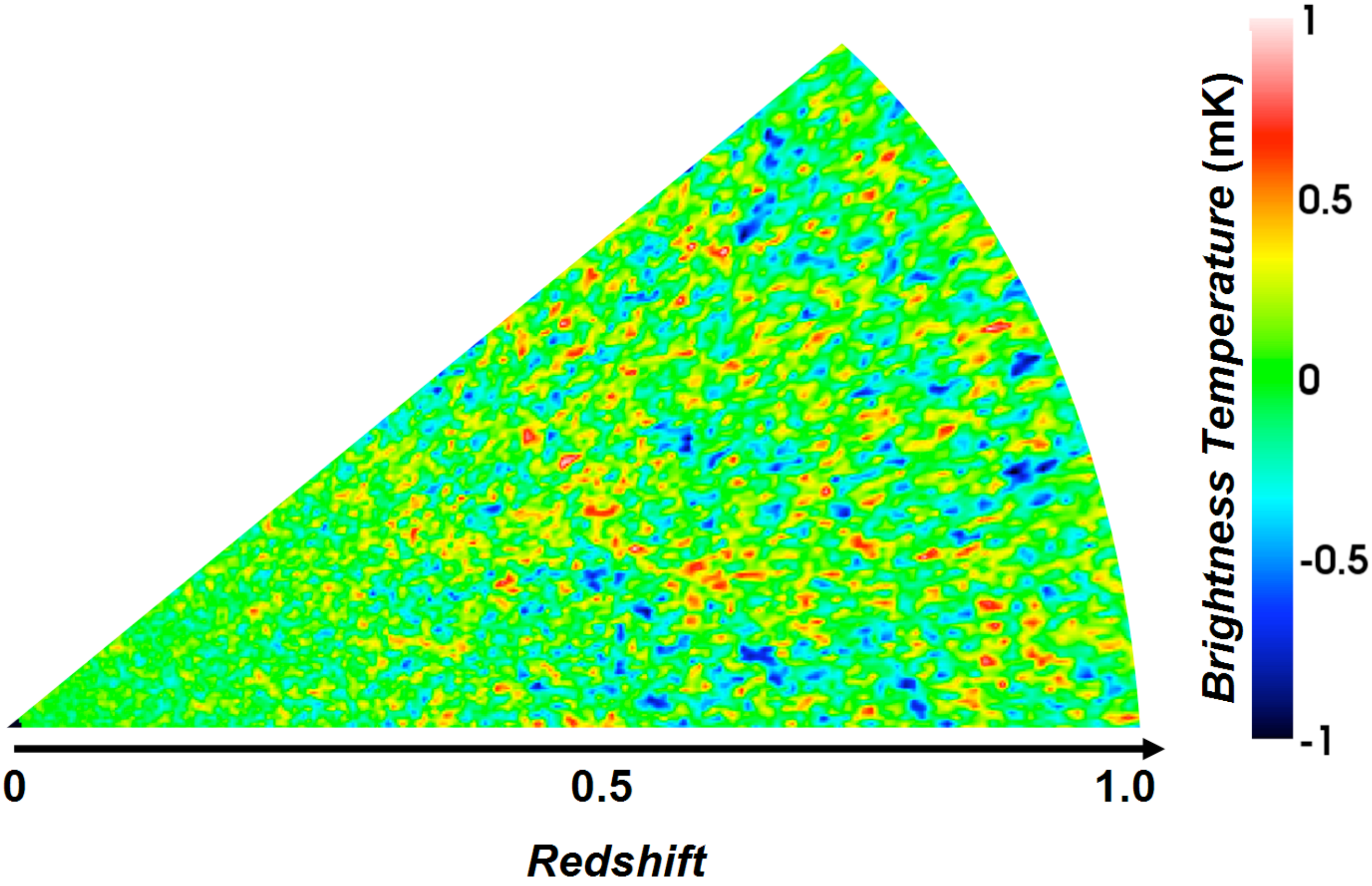

*Figure 2: Simulated fluctuations in the brightness temperature of 21cm emission from galaxies in a slice through the universe. The emission is smoothed over 8/h Mpc. The redshift, z, translates to frequency:* $\nu$=1.42GHz/(1+$z$). *Red indicates overdensity and blue underdensity.*

**Partnership between Developed and Developing Nations.**
Developing countries still have sites that offer very low levels of man-made radio interference, while at many radio telescope sites in developed countries this man-made interference severely impacts observations. Meanwhile, it is the developed countries that have the science and engineering infrastructure needed to design the next generation of radio telescope. It is natural then that partnerships between developing and developed countries be explored. This has the benefit of allowing expansion of outreach efforts in developing nations where improvement of educational outcomes can have dramatic positive social impact.

**Cost Factors.**
While the Decadal Survey Panel will rightly focus first on prioritizing science goals, cost factors must ultimately also be considered when setting priorities. We point out that the mapping-speed to cost ratio for an intensity mapping survey has improved dramatically over the last decade and should improve further during the teens. It is this improvement of cost efficiency will enable a new generation of instrument with mapping speed 1000 times that of existing instruments.

A few examples are needed to make this argument specific. First, costs for the sampling and signal processing hardware have been falling in a manner similar to other Information Technology costs. Several radio astronomy labs have developed sampling/DSP systems, which currently cost about $1500/ channel. The cost performance of DSP systems has followed a Moore's law trajectory during the 00s, and has reached the point that systems with 10,000 channels can now be considered. While it is not clear that Moore's Law cost reductions will continue apace to the end of the coming decade, at least several more factors of two improvement in performance/cost are likely. Furthermore, most current designs employ field-programmable technology. That's sensible for the current generation of prototype instrument, but substantial additional cost reductions are possible in the next decade, by creation of application-specific or custom-mask silicon devices. Second, at GHz frequencies, construction costs of the radiation collection component of the intensity-mapping system can be quite low. The 21CMA, a 21-cm transit telescope designed for 21-cm observations at $z \sim 10$, was built in China over the last few years for about $4M. This telescope has about 10,000 square meters of collecting area. 21CMA is an aperture-plane array, but fixed reflectors can also be inexpensive. For example, the Pittsburgh Cylinder Prototype telescope at Carnegie Mellon was built for less than $200/ sq meter.

The cost efficiency factors for GHz radio astronomy are right now at a tipping point. Current radio astronomy instruments use 1-20 receiver channels per aperture. In the coming decade the transition to thousands of channels will take place.

**Time Domain.**

The instruments being discussed for 21 cm intensity mapping offer dramatic advance of mapping speed combined with daily coverage of the entire sky. These instruments will therefore generate a rich set of data for synoptic studies. Pulsars, radio afterglows of gamma ray bursts, and variations of AGN emission are the main currently-known time variable radio sources. One can only guess what else might be uncovered using these new instruments.

**Conclusions.**

Optical astronomers have recognized the value of telescopes dedicated to sky surveys, in addition to user facilities. The radio astronomy community should consider this path as well. All-sky surveys can employ fixed, transit instruments. This allows large collecting areas at low cost and allows for telescope layouts that readily accommodate thousands of receiver channels. The mapping speed available with such instruments will exceed that available today by a factor 1000, offering extensive discovery potential. GHz radio astronomy stands out as an area of astronomy that, in the coming decade, offers potential for dramatic expansion of sensitivity.